# Structural, magnetic and transport properties of thin films of the Heusler alloy $Co_2MnSi$

L. J. Singh[a)] and Z. H. Barber

Department of Materials Science and Metallurgy, University of Cambridge, Pembroke Street, Cambridge CB2 3QZ, UK

Y. Miyoshi, Y. Bugoslavsky, W. R. Branford and L. F. Cohen

Blackett Laboratory, Imperial College, Prince Consort Road, London SW7 2AZ, UK

[a)]Electronic mail: ljl26@cam.ac.uk

**Abstract**

Thin films of $Co_2MnSi$ have been grown on a-plane sapphire substrates from three elemental targets by dc magnetron co-sputtering. These films are single phase, have a strong (110) texture and a saturation magnetization of 4.95 $\mu_B$/formula unit at 10 K. Films grown at the highest substrate temperature of 715 K showed the lowest resistivity (47 $\mu\Omega$cm at 4.2 K) and the lowest coercivity (18 Oe). The spin polarization of the transport current was found to be of the order of 54% as determined by point contact Andreev reflection spectroscopy. A decrease in saturation magnetization with decreasing film thickness and different transport behaviour in thinner films indicate a graded disorder in these films grown on non-lattice matched substrates.

The growing interest in Heusler alloys is due to their predicted high spin polarization (P) (i.e. dominance of carriers of one spin orientation at the Fermi energy). Such materials could prove to be a useful source of spin to inject into semiconductors. Some Heusler alloys have been predicted to be half-metallic ferromagnets (HMF), in which a band gap opens at the Fermi level ($E_F$) for one spin direction, leading to 100% spin polarized conduction electrons. There are two families of Heusler alloys: the half Heusler alloys which have compositions of the form XYZ and the full Heusler alloys, which have compositions of the form $X_2YZ$. Experimentally, the spin polarization of the transport current, $P_t$ of the HMF oxide $CrO_2$ has been measured by point contact Andreev reflection spectroscopy as 96%[1] but disappointingly for the half Heusler NiMnSb, $P_t$~58%.[2] It is important to identify the mechanisms that cause this half Heusler not to achieve its predicted 100% $P_t$ at the free surface. Three mechanisms have been identified; surface reconstruction and segregation,[3] and interatomic disorder.[4]

$Co_2MnSi$ is a full Heusler alloy that has been predicted to be a HMF.[5] It crystallizes in the $L2_1$ structure (space group Fm3m), which consists of four interpenetrating FCC sublattices.[6] This could be a very promising material for microelectronic applications: it is predicted to have a large energy gap in the minority band of ~0.4 eV[7] and has the highest Curie temperature amongst the known Heuslers of 985 K.[8] However, $Co_2MnSi$ (along with other Heuslers) can suffer from antisite disorder[6] because the Co and Mn atoms have similar atomic radii. It has been grown in thin film form from a single, stoichiometric target[9-11] but it has been reported[10] that this resulted in films that were deficient in Si, and this deviation from the ideal stoichiometry resulted in a lower value of the saturation magnetization ($M_s$) than in the bulk.

We have grown thin films of $Co_2MnSi$ by dc magnetron co-sputtering from three, elemental targets (as opposed to a single alloy target) in order to control the stoichiometry precisely, as well as to study off-stoichiometric compositions. In this letter we report on the transport properties and present the first transport spin polarization measurements carried out on thin films of this alloy. This study shows that optimized films have properties that are close to those of bulk single crystals.[12,13]

Films were deposited from three, elemental dc magnetron targets arranged in a triangular configuration, onto an array of a-plane sapphire substrates located directly below the targets on a Ta strip heater. The base pressure of the deposition chamber was $2 \times 10^{-9}$ Torr and the argon pressure during the film deposition was 24 mTorr. The target-substrate distance varied between 111 and 116 mm (measured from the centre of each target to the substrate position that gives the desired stoichiometry). This geometry enabled a range of compositions to be attained on the array of substrates, but small variation across each substrate (4×9 mm). The depositions were carried out at a range of substrate temperatures ($T_{sub}$) from 545 to 715 K, and a range of film thickness from 110 to 400 nm were deposited. Film thickness was determined by profilometry of a step formed on a masked substrate: the deposition rate was 0.10 nm/s. Film compositions were determined by energy dispersive X-ray analysis in a scanning electron microscope (SEM) with a precision of 1.5%.

Structural characterisation was performed by X-ray diffraction (XRD) and the XRD pattern of a 400 nm, stoichiometric film grown at $T_{sub}$ of 715 K is shown in Fig. 1. The film is single phase (within the resolution of the measurement) and all diffraction peaks can be attributed to the $Co_2MnSi$ $L2_1$ structure. The film is polycrystalline with a strong (110) texture. Magnetic measurements carried out in a vibrating sample magnetometer and a superconducting quantum interference device magnetometer from room temperature down to 10 K also indicate that the correct phase has been achieved. In stoichiometric films $M_s$ (determined from in-plane hysteresis loops at 10 K) was 1007 ± 50 emu/cc (4.95 ± 0.25 $\mu_B$/formula unit), which agrees well with the bulk value of 5.1 $\mu_B$/formula unit.[13]

Transport measurements were carried out from 4.2 to 295 K using a standard DC 4-point geometry. The dependence of $\rho_{4.2\ K}$ and room temperature coercivity ($H_c$) on $T_{sub}$ is shown in Fig. 2. These films are all stoichiometric and are 400 nm thick. $\rho_{4.2\ K}$ decreases with increasing $T_{sub}$ which is most likely due to an increase in grain size and hence a reduction in grain boundary scattering. As $T_{sub}$ increases from 545 to 715 K the surface features (which reflect the grain size) observed in a field emission gun SEM increase from about 70 to 130 nm, and $\rho_{4.2\ K}$ is halved (102 to 47 μΩcm).

For a stoichiometric film grown at $T_{sub}$ of 715 K, the ρ-T curve was fitted: in the temperature range 295 to 100 K, ρ decreased linearly with temperature. This differs from previous reports[14,15] in which a $T^n$ dependence is observed, where $1.3 < n < 1.5$. However, not all our films show linear behaviour and, as the compositions become more off-stoichiometric, the non-linearity becomes more pronounced. Below 100 K (for the stoichiometric film grown at 715 K) there is a change in the resistivity behaviour, and a $T^2 + T^{9/2}$ behaviour is observed. The $T^{9/2}$ term is attributed to two-magnon scattering, which is the first available magnetic scattering process for a HMF. The $T^2$ term results either from one magnon scattering or from electron-electron scattering. In a HMF, one magnon scattering cannot occur at low temperature because there are no states at $E_F$ into which the electron can be scattered. Therefore, if these films are truly HMF, then the $T^2$ term must arise from electron-electron scattering. Below about 20 K, ρ shows no temperature dependence, which is in agreement with the findings of Ambrose *et al.*[14] but not with other reports.[15-17] The different behaviours observed could be material dependent, but are also dependent upon fabrication parameters (different behaviour observed even for the same alloy), as crystallography and microstructure will affect the ρ-T behaviour.

$H_c$ also decreases with $T_{sub}$ as shown in fig.2, where an increase in $T_{sub}$ from 545 to 715 K results in a reduction in $H_c$ by a factor of three ($M_s$ is a constant in these films). The change in $H_c$ is crystallographic and/or microstructural in origin: both grain size and degree of (110) texturing increase as $T_{sub}$ increases. $H_c$ correlates strongly with the change in film texture (inset to fig. 2). As $T_{sub}$ increases, the (422) peak decreases relative to the (220) peak and $H_c$ decreases.

Fig. 3 shows the dependence of $\rho_{4.2\ K}$ and $M_s$ on film thickness, d for a series of stoichiometric films grown at a $T_{sub}$ of 602 K. $\rho_{4.2\ K}$ decreases with increasing film thickness. This can be attributed to the reduction in grain boundary scattering: surface feature size increases from about 20 to 70 nm as the film thickness increases from 110 to 300 nm. Thinner films also show a stronger (110) texture. $M_s$ decreases by about 25% as the film thickness decreases from 400 to 110 nm. This indicates that thinner films are disordered, due to the absence of lattice matching between the film and substrate.

A change in the ρ-T curves is also observed in the above series of films as the film thickness decreases: an upturn in the resistivity at low temperatures is present for the 164 nm film and becomes more pronounced in the 110 nm film (Fig. 4). This upturn has been observed in $Ni_2MnGe$ at 20 K by Lund *et al.*[16] and in $Co_2MnSi$ at 40 K by Geiersbach *et al.*[10] We have also observed this phenomenon in off-stoichiometric films.[18] The upturn is interpreted as weak localization[16] due to electron-electron interaction effects, arising in this case from disordered material, consistent with the decrease in $M_{s.}$

$P_t$ was measured by point contact Andreev reflection[2,19] between Nb cut tips and the Heusler film at 4.2 K. The normalised conductance vs voltage curves were fitted to the Mazin model in the ballistic regime,[19] to determine $P_t$ and the effective tunnel barrier height, Z. The barrier may arise due to the combined effects of surface contamination or disorder, Fermi velocity mismatch in the two materials and interface scattering. Experimentally, Z can be varied to a certain extent by changing the tip pressure (Z=0 corresponds to a clean interface and is obtained at higher pressure). The inset to Fig. 4 shows the variation of $P_t$ with Z for the 400 nm film grown at 715 K (ρ value shown in Fig. 2). At Z=0, $P_t$~54 ± 3%, consistent with measurements on bulk single crystals of $Co_2MnSi$.[12,13] However, for the thinnest film (110 nm) it was only possible to obtain contacts with high values of Z. One data point for this film is shown for comparison. The larger error bars reflect the fact that in this Z range the fitting is less reliable. Nevertheless the polarization is depressed in the thinner film within the error of our measurement. As the surface quality of all the films is similar, the non-vanishing barrier in the 110 nm film suggests that the intrinsic properties of this film are different from the others – in line with the transport and magnetization data taken on this film. Further investigation is needed to gain a detailed understanding of the processes involved in Andreev reflection in the thinnest film.

In summary, we have grown $Co_2MnSi$ from three elemental targets onto a-plane sapphire substrates by dc magnetron co-sputtering and have obtained highly textured, single phase films without the use of a seed layer.[10,11] Stoichiometric, 400 nm films had the bulk value of $M_s$ and films grown at the highest $T_{sub}$ showed the lowest ρ and $H_c$. $P_t$ of a 400 nm film grown at this deposition temperature was 54%. A decrease in $M_s$ with decreasing film thickness and the different transport behaviour of thinner films indicates a graded disorder, which is significant even at a thickness of 110 nm. Increased interface scattering in disordered magnetic films is an interesting independent observation. By growing on GaAs, which has a similar lattice parameter to $Co_2MnSi$ it is expected that this disordered region will be confined to the first few atomic layers.

This work was supported by the Engineering and Physical Sciences Research Council, UK.

List of Figures

Fig. 1. XRD pattern of a stoichiometric, 400 nm film. (The sharp lines indexed with a C, are W contamination lines from the X-ray tube and the weak, broad peaks at 20 and 60 °2θ were also observed on a bare substrate).

Fig. 2. The effect of $T_{sub}$ on $\rho_{4.2\ K}$ and room temperature $H_c$ for a series of 400 nm, stoichiometric films. The inset shows the dependence of room temperature $H_c$ on the (422):(220) XRD peak intensity ratio.

Fig. 3. The effect of film thickness on $\rho_{4.2\ K}$ and $M_s$ for a series of stoichiometric films grown at 602 K.

Fig. 4. ρ-T curves for stoichiometric films of different thickness, grown at the same deposition temperature of 602 K. For display purposes the curves have been translated along the y-axis to coincide at 295 K. The vertical axis corresponds to the 110 nm film. $\rho_{295\ K}$ are 89, 117, and 124 μΩcm and $\rho_{4.2\ K}$ are 66, 96, and 106 μΩcm for the 300, 164 and 110 nm films respectively. The inset shows the variation in $P_t$ with Z for a stoichiometric, 400 nm film grown at 715 K (filled squares) and the 110 nm film (open square) shown in the main part of the figure.

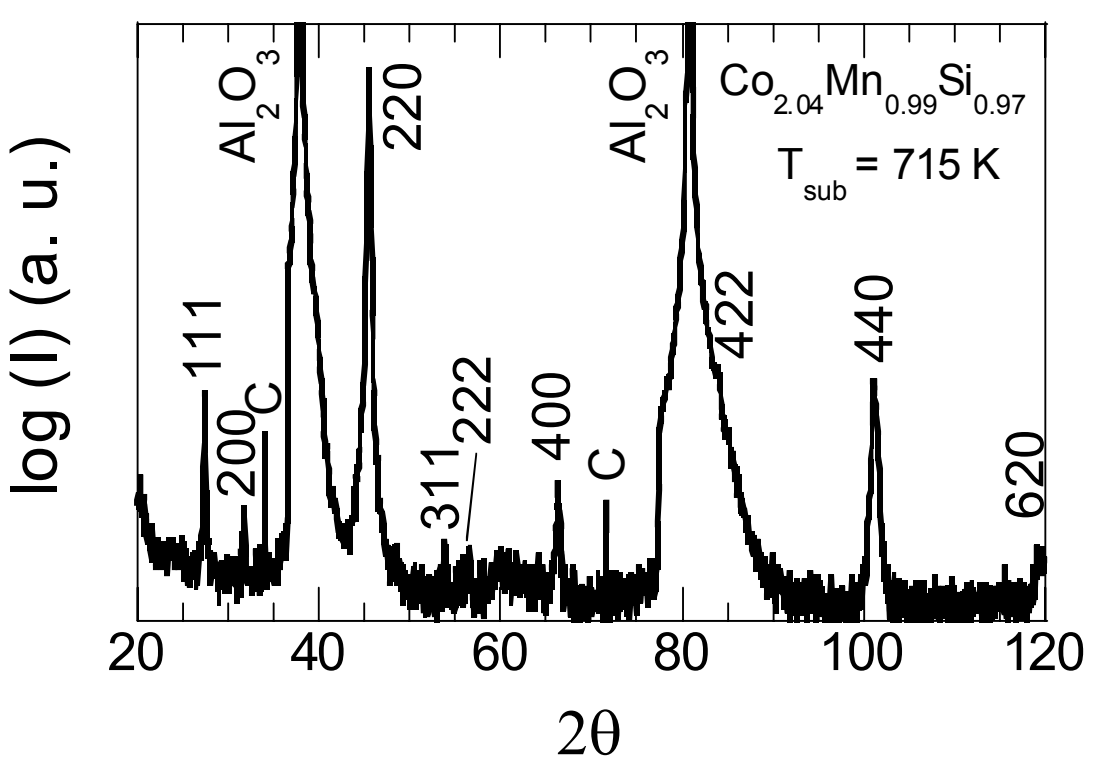

Fig. 1.

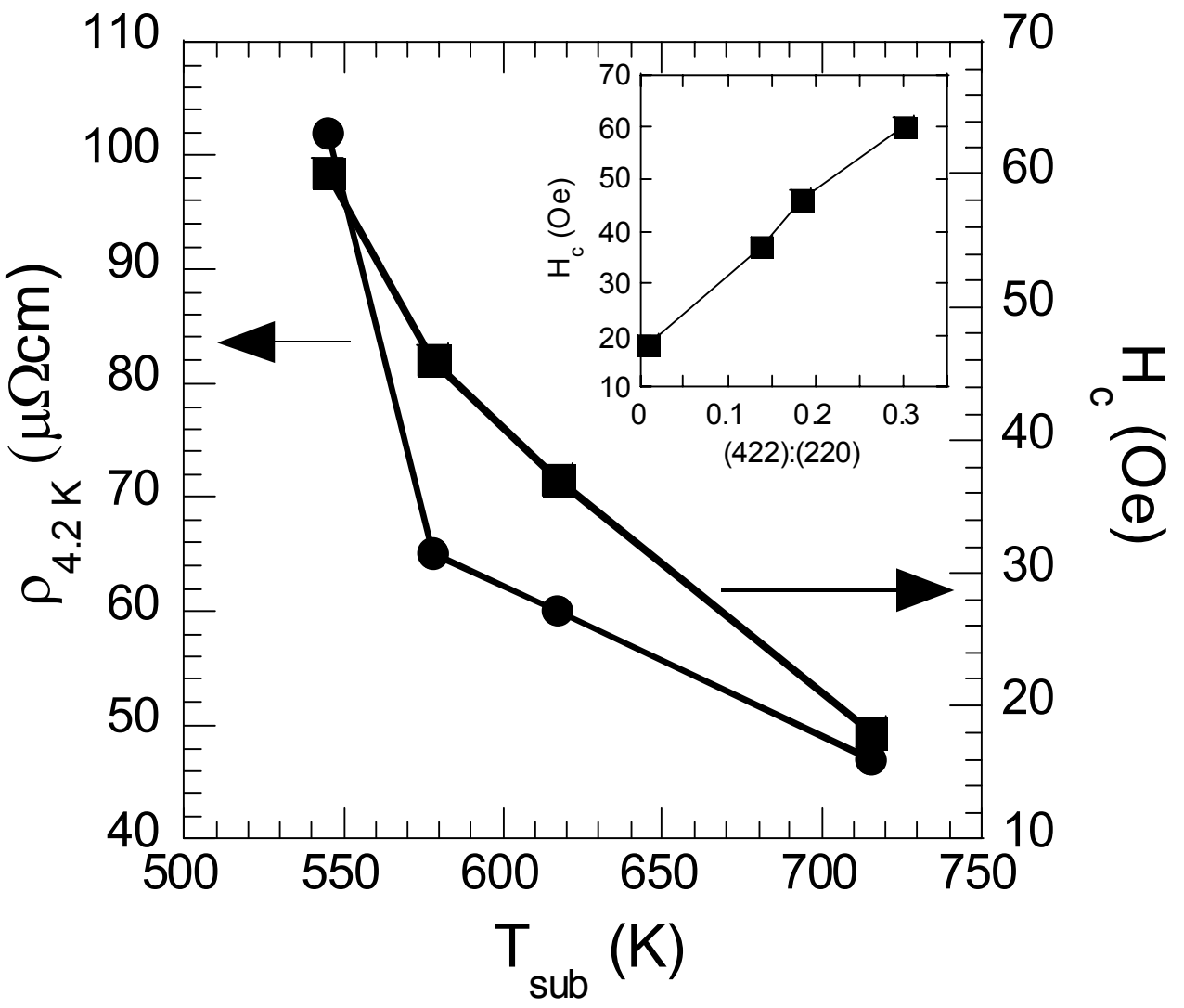

Fig. 2.

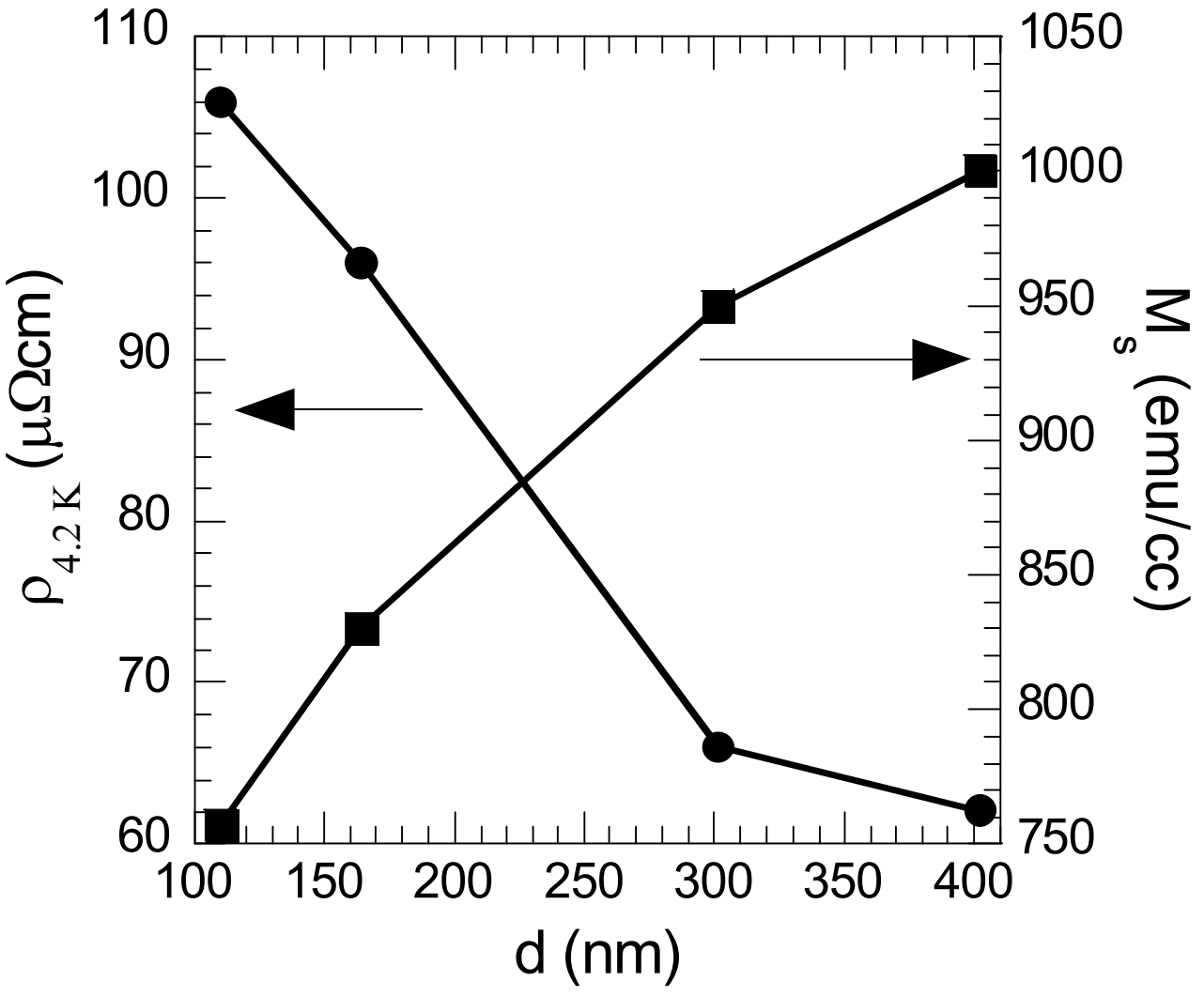

Fig. 3.

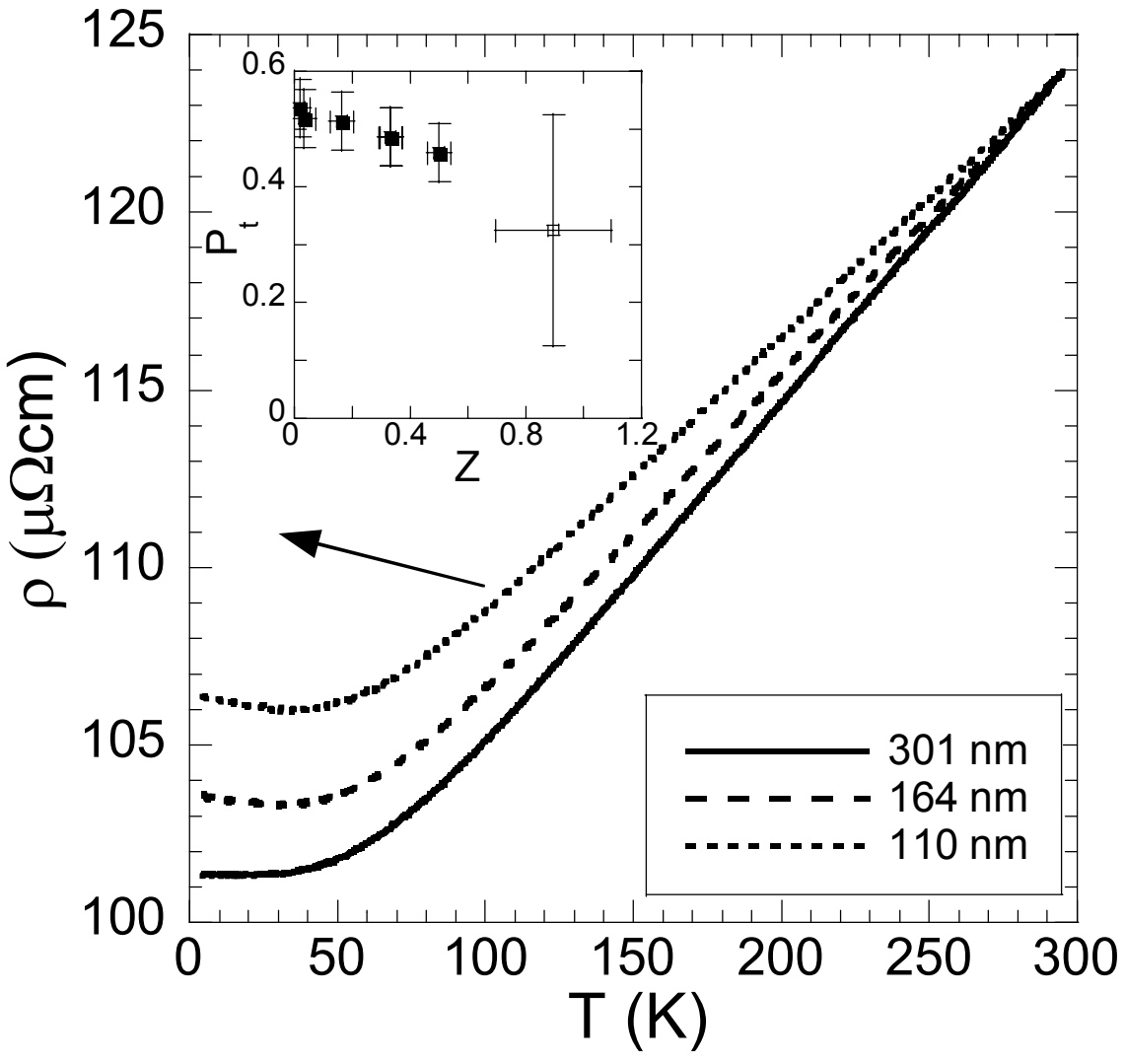

Fig. 4.